\documentclass[referee]{jfm}
\usepackage{color}
\usepackage{amstext}
\usepackage{amssymb}
\usepackage{graphicx}
\usepackage{amsmath}

\newcommand\curl{\mathrm{curl} \,}

\newcommand\const{\mathrm{const}}
\newcommand\Div{\mathrm{div}\,}

\newcommand\vV{\boldsymbol{V}}

\newcommand\vf{\boldsymbol{f}}
\newcommand\vh{\boldsymbol{h}}

\newcommand\vu{\boldsymbol{u}}

\newcommand\vx{\boldsymbol{x}}

\newcommand\vg{\boldsymbol{g}}

\newcommand\vomega{\boldsymbol{\omega}}

\newcommand\vlambda{\boldsymbol{\lambda}}

\begin{document}

{\title[ Acoustic-drift equation ] {Acoustic-drift equation }}

\author[V. A. Vladimirov]
{V.\ns A.\ns V\ls l\ls a\ls d\ls i\ls m\ls i\ls r\ls o\ls v}

\affiliation{Dept of Mathematics, University of York, Heslington, York, YO10 5DD, UK}

\pubyear{2010} \volume{xx} \pagerange{xx-xx}
\date{Sept 12th 2011}

\setcounter{page}{1}\maketitle \thispagestyle{empty}

\begin{abstract}

The aim of this paper is to derive a new equation (the
\emph{acoustic-drift equation} (ADE)) describing the generation of a flow by an acoustic wave.
We consider acoustic waves of perfect barotropic gas as the zero-order solution and derive the equation for
the averaged flow of the first order. The used small parameter of our asymptotic study is dimensionless
inverse frequency, and the leading term for a velocity field is chosen to be a purely oscillating acoustic
field. The employed mathematical approach combines the two-timing method and the notion of a distinguished
limit. The properties of commutators are used to simplify calculations. The derived averaged equation is
similar to the original vorticity equation, where the Reynolds stresses has been transformed to an additional
advection with the drift velocity. Hence ADE can be seen as a compressible version of the Craik-Leibovich
equation. In particular, ADE shows that if the averaged vorticity is absent in the initial data, then it will
always be equal to zero. This property confirms that the acoustic streaming is of viscous nature. At the same
time ADE can be useful for the description of secondary motions which take place for `strong' acoustic
streaming flow.

\end{abstract}

\section{Introduction \label{sect01}}

Acoustic streaming represents a classical topic of fluid dynamics, see \emph{e.g.}
\cite{Riley,Nyborg0,Lighthill0,Lighthill}. It usually describes the averaged flow generated by a given acoustic wave.
The related asymptotic analysis is usually based on the Navier-Stokes equations and goes up to the
second-order (in amplitude) terms, while the terms of the third and higher orders appear as physically
irrelevant. In contrary to this classical approach we build a rigorous asymptotic procedure based on the
Euler equations where the first-order averaged equations (we call it an acoustic-drift equation (ADE))
appears from the treatment of the third-order approximation (when the primary acoustic wave is treated as the
zero order).

Mathematically, we suggest that velocity, density, and pressure oscillate with frequency higher than inverse
characteristic times of all other co-existing motions. It can be expressed as $\sigma\gg 1$, where $\sigma$
is  dimensionless frequency of oscillations. To derive the averaged equation we employ the two-timing method
in the form used by \cite{VladimirovMHD}. The resulted averaged vorticity equation is similar to the original
equation, but surprisingly a drift velocity (instead of Reynolds stresses)  plays a part of an additional
advection velocity. The derived equation for the averaged first-order vorticity (ADE) formally coincides with
the Craik-Leibovich equations for an incompressible fluid (\cite{CraikLeib, Leibovich, Craik0,
VladimirovMHD}), while the difference in our case appears as the prescribed divergence of velocity. An
additional incentive for our research is to make all the calculations and derivations elementary and free of
any physical and mathematical assumptions (except the most common ones, such as the existence of
differentiable solutions). We use only Eulerian description and Eulerian averaging operation, since they are
the most transparent and allow us to avoid any unnecessary difficulties.

\section{General setting}

\subsection{Formulation of problem}

The governing equation for  an  inviscid compressible barotropic fluid (gas) with velocity field $\vu^*$,
density $\rho^*$, vorticity $\vomega^*\equiv\curl^*\vu^*$ and pressure $p^*$ is taken in the vorticity form
\begin{eqnarray}\label{exact-1}
&&{\partial\vomega^*}/{\partial {t}^*}+[\vomega^*,\vu^*]^*+\vomega^* \Div\vu^*=0,\quad
\text{in}\quad \mathcal{D}^*\\
&&{\partial\rho^*}/{\partial {t}^*}+(\vu^*\cdot\nabla)\rho^*+\rho^*\Div \vu^*=0=0\nonumber
\end{eqnarray}
where asterisks mark dimensional variables and operations, ${t}^*$-time,
$\vx^*=(x_1^*,x_2^*,x_3^*)$-cartesian coordinates, and $[\,\cdot\, ,
\cdot\,]^*$ stands for the dimensional commutator (\ref{commutr}). The term with pressure has been
excluded since the fluid is barotropic $p^*=p^*(\rho^*)$. In this paper we deal with the transformations of
equations, hence the form of flow domain $\mathcal{D}^*$ and particular boundary conditions can be specified
at the later stages. The related equation for the ratio $\vlambda^*\equiv\vomega^*/\rho^*$ is
\begin{eqnarray}\label{exact-lambda}
&&{\partial\vlambda^*}/{\partial {t}^*}+[\vlambda^*,\vu^*]^*=0,\quad
\text{in}\quad \mathcal{D}^*
\end{eqnarray}
We accept that the considered class of (unknown) oscillatory solutions $\vu^*,\rho^*$ possesses
characteristic scales of velocity $U$, length $L$, density $\rho_0^*$ and high frequency $\sigma^*$
\begin{eqnarray}
&& U,\quad L,\quad \rho_0^*,\quad \sigma^*\gg 1/T;\quad T\equiv L/U
\label{scales-list}
\end{eqnarray}
where $\rho_0^*$ is reference density (at the state of rest), $T$ is a dependent time-scale. The
dimensionless variables and frequency are
\begin{eqnarray}
&& \vx\equiv\vx^*/L,\quad t\equiv t/T,
\quad\vu\equiv\vu^*/U,\quad \rho\equiv \rho^*/\rho_0^*,
\quad\sigma\equiv\sigma^*T\gg 1
\label{scales}
\end{eqnarray}

\subsection{Notations}

The variables $\vx=(x_1,x_2,x_3)$, $t$, $s$, and $\tau$  in the text below serve as  dimensionless cartesian
coordinates, physical time, slow time, and fast time. The used definitions, notations, and properties are:

\noindent
(i) A dimensionless function $f=f(\vx,s,\tau)$ belongs to the class $\mathbb{O}(1)$ if $f={O}(1)$ and all
partial $\vx$-, $s$-, and $\tau$-derivatives of $f$ (required for our consideration) are also ${O}(1)$. In
this paper all small parameters appear as explicit multipliers, while all functions always belong to
$\mathbb{O}(1)$-class.

\noindent
(ii) The class $\mathbb{H}$ of \emph{hat-functions}  (or oscillating functions with the non-zero mean) is
defined as
\begin{eqnarray}
\widehat{f}\in \mathbb{H}:\quad
\widehat{f}(\vx, s, \tau)=\widehat{f}(\vx,s,\tau+2\pi)\label{tilde-func-def}
\end{eqnarray}
where the $\tau$-dependence is always $2\pi$-periodic; the dependencies on $\vx$ and $s$ are not specified.

\noindent
(iii) The subscripts $t$, $\tau$, and $s$ denote the related partial derivatives.

\noindent
(iv) For an arbitrary $\widehat{f}\in \mathbb{H}$ the \emph{averaging operation} is
\begin{eqnarray}
\langle {\widehat{f}}\,\rangle \equiv \frac{1}{2\pi}\int_{\tau_0}^{\tau_0+2\pi}
\widehat{f}(\vx, s, \tau)\,d\tau,\qquad\forall\ \tau_0\label{oper-1}
\end{eqnarray}
where  during the $\tau$-integration we keep $s=\const$ and $\langle {\widehat{f}}\rangle$ does not depend on
$\tau_0$.

\noindent
(v)  The class $\mathbb{T}$ of \emph{tilde-functions} (or purely oscillating functions, or fluctuations) is
such that
\begin{eqnarray}
\widetilde f\in \mathbb{T}:\quad
\widetilde f(\vx, s, \tau)=\widetilde f(\vx,s,\tau+2\pi),\quad\text{with}\quad
\langle \widetilde f \rangle =0,\label{oper-2}
\end{eqnarray}
they represent a special case of hat-functions with the zero average. The class $\mathbb{B}$ of
\emph{bar-functions} (or mean-functions) is defined as
\begin{eqnarray}
\overline{f}\in \mathbb{B}:\quad  \overline{f}_{\tau}\equiv 0,\quad
\overline{f}(\vx, s)=\langle\overline f(\vx,s)\rangle
 \label{oper-3}
\end{eqnarray}

\noindent
(vi) We introduce \emph{the tilde-integral} (or the fluctuating part of an integral of a fluctuating
function) $\widetilde{f}^{\tau}$ as
\begin{eqnarray}
&&\widetilde{f}^{\tau}\equiv\int_0^\tau \widetilde{f}(\vx,s,\eta)\,d\eta
-\frac{1}{2\pi}\int_0^{2\pi}\Bigl(\int_0^\mu
\widetilde{f}(\vx,s,\eta)\,d\eta\Bigr)\,d\mu\label{oper-7}
\end{eqnarray}
The tilde-integration is inverse to $\tau$-differentiation
$(\widetilde{f}^{\tau})_{\tau}=(\widetilde{f}_{\tau})^{\tau}=\widetilde{f}$; the proof is omitted. The
$\tau$-derivative of a tilde-function always represents a tilde-function. However the $\tau$-integration of a
tilde-function can produce a hat-function. \{For example: let us take
$\widetilde\phi=\overline\phi_0\sin\tau$ where $\overline\phi_0$ is an arbitrary bar-function: one can see
that $\langle\widetilde\phi\,\rangle\equiv 0$, however
$\langle\int_0^\tau\widetilde\phi(\vx,s,\rho)d\rho\rangle=\overline\phi_0\neq 0$, unless
$\overline\phi_0\equiv 0$.\} Formula (\ref{oper-7}) keeps the result of integration inside the
$\mathbb{T}$-class.

\noindent
(vii) The unique solution of a PDE inside the tilde-class
\begin{eqnarray}
\widetilde{f}_\tau\equiv\partial \widetilde{f}/\partial\tau=0\quad \Rightarrow\quad
\widetilde{f}\equiv 0\label{f-tilde=0}
\end{eqnarray}
follows from (\ref{oper-7}). One can also see that $\widehat{f}_\tau=\widetilde{f}_\tau, \quad
\langle\widehat{f}_\tau\rangle=\langle\widetilde{f}_\tau\rangle=0$.

\noindent
(viii) The commutator of two vector-fields $\vf$ and $\vg$ is
\begin{eqnarray}
&& [\vf,\vg]\equiv(\vg\cdot\nabla)\vf-(\vf\cdot\nabla)\vg\label{commutr}
\end{eqnarray}
The commutator is antisymmetric and satisfies Jacobi's identity for any vector-fields $\vf$, $\vg$, $\vh$:
\begin{eqnarray}
&&[\vf,\vg]=-[\vg,\vf],\quad [\vf,[\vg,\vh]]+ [\vh,[\vf,\vg]]+[\vg,[\vh,\vf]]=0 \label{oper-13}
\end{eqnarray}
As the average operation (\ref{oper-1}) is proportional to the integration over $\tau$, the integration by
parts yields
\begin{eqnarray}
&&\langle[\widetilde{\vf},\widetilde{\vg}_\tau]\rangle=-\langle[\widetilde{\vf}_\tau,\widetilde{\vg}]\rangle=-
\langle[\widetilde{\vf}_\tau, \widehat{\vg}]\rangle,\
\langle[\widetilde{\vf},\widetilde{\vg}^\tau]\rangle=-\langle[\widetilde{\vf}^\tau,\widetilde{\vg}]\rangle=-
\langle[\widetilde{\vf}^\tau, \widehat{\vg}]\rangle
\label{oper-15}
\end{eqnarray}
For any tilde-function $\widetilde{\vf}$ and bar-function $\overline{\vg}$ (\ref{oper-13}),(\ref{oper-15})
give
\begin{eqnarray}
\langle[\widetilde{\vf},[\overline{\vg},\widetilde{\vf}^\tau]]\rangle=[\overline{\vg},\overline{\vV}]\quad
\text{where}\quad\overline{\vV}\equiv\langle[\widetilde{\vf},\widetilde{\vf}^\tau]\rangle/2\label{lem1}
\end{eqnarray}

\section{Asymptotic procedure}

\subsection{Two-timing setting and distinguished limit}

We assume that a flow has its own intrinsic slow-time scale $T_{\text{slow}}$ (which can be different from
$T$) and consider solutions of (\ref{exact-1}) in the form of hat-functions (\ref{tilde-func-def})
\begin{eqnarray}
&& \vu^*=U \widehat{\vu}(\vx, s, \tau),\quad \rho^*=\rho_0 \widehat{\rho}(\vx, s, \tau);\quad  \text{with}\,
\
\tau\equiv\sigma {t},\, s\equiv\Omega\, {t},
\, \Omega\equiv T/T_{\text{slow}}\label{exact-2}
\end{eqnarray}
Then the use of the chain rule and the transformation to dimensionless variables give
\begin{eqnarray}
&&\left(\frac{\partial}{\partial\tau}+
\frac{\Omega}{\sigma}\frac{\partial}{\partial s}\right)\widehat{\vomega}+
\frac{1}{\sigma}[\widehat{\vomega},\widehat{\vu}]+\frac{1}{\sigma}\vomega\Div\vu
= 0\label{exact-6-h}\\
&&\left(\frac{\partial}{\partial\tau}+
\frac{\Omega}{\sigma}\frac{\partial}{\partial s}\right)\widehat{\rho}
+\frac{1}{\sigma}(\vu\cdot\nabla)\rho+\frac{1}{\sigma}\rho\Div \vu=0=0\nonumber
\end{eqnarray}
In order to keep variable $s$ `slow' in comparison with $\tau$ we have to accept that $\Omega/\sigma\ll 1$.
Then eqn.(\ref{exact-6-h}) contains two independent small parameters:
\begin{eqnarray}
\varepsilon\equiv \frac{1}{T\sigma^*}=\frac{1}{\sigma},\quad \varepsilon_1\equiv
\frac{1}{T_\text{slow}\sigma^*}\equiv\frac{\Omega}{\sigma}\label{small-pars-h}
\end{eqnarray}
Here we must make an auxiliary (but technically essential) assumption: after the use of the chain rule
(\ref{exact-6-h}) variables $s$ and $\tau$  are considered to be \emph{mutually independent}:
\begin{eqnarray}
&& \tau,\qquad s \qquad -\text{independent variables}\label{tau-s-ind}
\end{eqnarray}
Following the same distinguished limit argument as in (\cite{VladimirovMHD}) we take
\begin{eqnarray}
\Omega(\sigma)=1/\sigma:\quad\tau=\sigma t,\quad s=t/\sigma\label{tau-s-h}
\end{eqnarray}
In this paper we will show that the first three successive approximations (with $s$ (\ref{tau-s-h})) do
produce a valid asymptotic solution. The proof of other properties of the distinguished limit are similar to
the one considered in \cite{Vladimirov, VladimirovX2}; this proof is beyond the style and scope of this paper
and is omitted.

Hence the governing equations are
\begin{eqnarray}
&&\widehat{\vomega}_\tau+
\varepsilon[\widehat{\vomega},\widehat{\vu}]+\varepsilon \vomega\Div\vu
+\varepsilon^2\widehat{\vomega}_s= 0\label{exact-6eps}\\
&&\widehat{\rho}_\tau+
\varepsilon(\widehat{\vu}\cdot\nabla)\widehat{\rho}+\varepsilon \widehat{\rho}\Div\vu+\varepsilon^2\widehat{\rho}_s= 0
\label{exact-6eps-h}\\
&&\widehat{\vlambda}_\tau+\varepsilon[\widehat{\vlambda},\widehat{\vu}] +\varepsilon^2\widehat{\vlambda}_s=
0\label{exact-6eps-lambda}
\end{eqnarray}
where $\varepsilon\equiv 1/\sigma\to 0$. Let us look for the solutions of
(\ref{exact-6eps})-(\ref{exact-6eps-lambda}) in the form of regular series
\begin{eqnarray}
&&(\widehat{\vomega},\widehat{\vu},\widehat{\rho},\widehat{\vlambda})=\sum_{k=0}^\infty\varepsilon^k
(\widehat{\vomega}_k,\widehat{\vu}_k,\widehat{\rho}_k,\widehat{\vlambda}_k);\
\widehat{\vomega}_k,\widehat{\vu}_k,\widehat{\rho}_k,\widehat{\vlambda}_k \in \mathbb{H}\cap\mathbb{O}(1),\  k=0,1,2,\dots
\label{basic-4aa-0-h}
\end{eqnarray}
We study the class of flows which represents acoustic waves (and the related streaming flows) at the state of
rest:
\begin{eqnarray}
&& \overline{\vu}_0\equiv 0,\quad \overline{\rho}_0\equiv 1\label{mean0}
\end{eqnarray}
which implies that
\begin{eqnarray}
&& \overline{\vomega}_0\equiv 0,\quad \overline{\vlambda}_0\equiv 0\label{mean01}
\end{eqnarray}
The considered asymptotic series for velocity and density are
\begin{eqnarray}
&& \widehat{\vu}=\widetilde{\vu}_0+\varepsilon \widehat{\vu}_1+\varepsilon^2 \widehat{\vu}_2+\dots\label{vel-dens-ass}\\
&& \widehat{\rho}=1+\varepsilon \widetilde{\rho}_1+\varepsilon^2 \widehat{\rho}_2+\dots\nonumber
\end{eqnarray}
where $\widetilde{\vu}_0$ and $\widetilde{\rho}_1$ satisfies the classical linear acoustic equations that
easily follow after the substitution of (\ref{vel-dens-ass}) into the Euler's equations, which are related to
(\ref{exact-6eps}),(\ref{exact-6eps-h}) but not presented here for brevity. An important fact for Euler's
equations is: the application of $\curl^{-1}$ to (\ref{exact-6eps}) should produce a gradient of a function
(enthalpy) which must be $O(1)$ if we wish to build a valid asymptotic procedure; this order of enthalpy
requires an appropriate scale (in terms of $\varepsilon$) for pressure. In fact, all the derivation of ADE
can be carried ahead in terms of velocity field for the Euler equations, however in this paper we exploit the
simplification of calculations for the vorticity equation. It means that the fields
\begin{eqnarray}
&&\widetilde{\vu}_0= \nabla\phi_0(\vx,t), \quad \widetilde{\rho}_1=\widetilde{\rho}_1(\vx,t)\label{sol-00}
\end{eqnarray}
everywhere below are considered as a given solution of the classical inviscid linear acoustic equations.

\subsection{Successive approximations \label{sect04}}

The substitution of (\ref{basic-4aa-0-h})-(\ref{sol-00}) into (\ref{exact-6eps})-(\ref{exact-6eps-lambda})
produces the equations of successive approximations. Let us first consider the equation for
$\widehat{\vlambda}$ (\ref{exact-6eps-lambda}):

The zero-order equation is $\widehat{\vlambda}_{0\tau}=0$. Its unique solution (\ref{f-tilde=0}) is
$\widetilde{\vlambda}_{0}\equiv 0$ and $\widetilde{\vomega}_{0}\equiv 0$. Taking into account (\ref{mean01})
we can write
\begin{eqnarray}
&&\widehat{\vlambda}_0\equiv 0, \quad \widehat{\vomega}_0\equiv 0\label{sol-0}
\end{eqnarray}
which means that the zero-order flow  is potential and  purely oscillating. Eqn (\ref{sol-0}) leads to the
similar first-order approximation of $\widehat{\vlambda}_{1\tau}=0$ which have the unique solution
\begin{eqnarray}
&&\widetilde{\vlambda}_1\equiv 0,,\quad\overline{\vlambda}_1=\boxed{?}\label{sol-1}
\end{eqnarray}
where a mean function remains undetermined. The second-order equations that take into account
(\ref{mean01}),(\ref{sol-0}),(\ref{sol-1}) are
$\widetilde{\vlambda}_{2\tau}+[\overline{\vlambda}_1,\widetilde{\vu}_0]=0$ which after  tilde-integration
(\ref{oper-7}) yield
\begin{eqnarray}
&&\widetilde{\vlambda}_2=[\widetilde{\vu}_0^\tau,\overline{\vlambda}_1],\quad
\overline{\vlambda}_2=\boxed{?}\label{sol-2}
\end{eqnarray}
The third-order equation that takes into account (\ref{mean01}),(\ref{sol-0}),(\ref{sol-1}) is
\begin{eqnarray}
&&\widetilde{\vlambda}_{3\tau}+\overline{\vlambda}_{1s}+[\widehat{\vlambda}_2,\widetilde{\vu}_0]+
[\overline{\vlambda}_1,\widehat{\vu}_1]=0\nonumber
\label{eqn-3}
\end{eqnarray}
The bar-part (\ref{oper-1}) (or the averaged part) of this equation is
\begin{eqnarray}
&&\overline{\vlambda}_{1s}+ [\overline{\vlambda}_1,\overline{\vu}_1]+
\langle[\widetilde{\vlambda}_2,\widetilde{\vu}_0]\rangle=0\label{eqn-3-bar}
\end{eqnarray}
which can be transformed with the use of (\ref{sol-2}) and (\ref{lem1}) into the form
\begin{eqnarray}
&&\overline{\vlambda}_{1s}+ [\overline{\vlambda}_1,\overline{\vu}_1+\overline{\vV}_0]=0
\label{eqn-3-bar1-1}
\end{eqnarray}
where the drift velocity is
\begin{eqnarray}
&&\overline{\vV}_0\equiv\langle[\widetilde{\vu}_0,\widetilde{\vu}_0^\tau]\rangle/2\label{drift-vel}
\end{eqnarray}
where  $\widetilde{\vu}_0$ is given by (\ref{sol-00}). Taking into account that
$\overline{\vlambda}_1=\overline{\vomega}_1/\rho_0$ (\ref{drift-vel}) can be rewritten as
\begin{eqnarray}
&&\overline{\vomega}_{1s}+ [\overline{\vomega}_1,\overline{\vu}_1+\overline{\vV}_0]=0
\label{eqn-3-bar1-1a}
\end{eqnarray}
which is the same as
\begin{eqnarray}
&&\overline{\vomega}_{1s}+ \nabla\times(\overline{\vomega}_1\times(\overline{\vu}_1+\overline{\vV}_0))=
\overline{\vomega}_1\Div(\overline{\vu}_1+\overline{\vV}_0)
\label{eqn-3-bar1-1aa}
\end{eqnarray}
The next step of our consideration is to repeat a similar (but analytically simpler) asymptotic procedure for
the density equation (\ref{exact-6eps-h}). Omitting the calculations we state that it yields
\begin{eqnarray}
&&\Div(\overline{\vu}_1+\overline{\vV}_0)=0
\label{eqn-3-bar1-1aaa}
\end{eqnarray}
and the equation (\ref{eqn-3-bar1-1aa}) is reduced to
\begin{eqnarray}
&&\overline{\vomega}_{1s}+ \nabla\times(\overline{\vomega}_1\times(\overline{\vu}_1+\overline{\vV}_0))=0
\label{eqn-3-bar1-1aaaa}
\end{eqnarray}
Taking $\curl^{-1}$ of this equation yields
\begin{eqnarray}
&&\overline{\vu}_{1s}+ \overline{\vomega}_1\times\overline{\vu}_1  +
\overline{\vomega}_1\times\overline{\vV}_0=-\nabla\overline{\Pi}_1
\label{eqn-3-bar1-vel}
\end{eqnarray}
where $\overline{\Pi}_1$ is modified pressure. The component version of this equation (where we drop the
subscripts `1' of the first approximation and `0' for a drift, use the summation convention, and introduce
different modified pressure $\overline{p}$) is:
\begin{eqnarray}
&&\overline{u}_{is}+ (\overline{u}_k+\overline{V}_k)\frac{\partial \overline{u}_i}{\partial
x_k}=-\frac{\partial\overline{p}}{\partial x_i} -\frac{\partial \overline{V}_k}{\partial
x_i}\overline{u}_k,\quad \frac{\partial \overline{u}_k}{\partial x_k}=-\frac{\partial
\overline{V}_k}{\partial x_k}
\label{eqn-3-bar1-vel-comp}
\end{eqnarray}
New equations for a streaming flow (\ref{eqn-3-bar1-1aaa})-(\ref{eqn-3-bar1-vel-comp}) (which we call ADE)
represent the main result of this paper.

\section{Discussion}

1. One can see that ADE (\ref{eqn-3-bar1-vel-comp}) represents an  Euler's equation `modified' by  additional
`advection term' and  `linear friction term'
\begin{eqnarray}
&& -\frac{\partial \overline{V}_k}{\partial x_i}\overline{u}_k\nonumber
\label{fric-term}
\end{eqnarray}
This `effective friction' is linear in velocity $\overline{\vu}$ and can be positive or negative. This term
is important for the zero-order solutions (\ref{sol-00}) with the amplitude of acoustic wave changing in
space, similar to described in
\cite{Lighthill0, Nyborg0}. Indeed, with the decreasing of the amplitude $|\widetilde{\vu}_0|$
of an acoustic wave (\ref{sol-00}) with the distance (say, with the increasing the distance from the source
of sound), the value $|\overline{V}_0|$ (\ref{drift-vel}) is also decreasing. Therefore the `friction term'
become positive (which corresponds to `anti-friction' or `negative friction'); from the physical viewpoint it
should generate the acoustic streaming. This argument formally repeats the treatment of
\cite{Lighthill0,Nyborg0}. However returning to the averaged vorticity equation (\ref{eqn-3-bar1-1aaaa}),
one can see that if the average vorticity is zero at the initial data, then it will always be equal to zero.
It means that the presence of a variable in space amplitude is not sufficient for the generation of acoustic
streaming. This conclusion confirms the classical idea about essentially viscous nature of acoustic
streaming. At the same time ADE contradicts to the estimations of acoustic streaming obtained without taking
viscosity into consideration, see \cite{Nyborg0} p.214, where a prescribed attenuated plane wave is
introduced `by hands' into an inviscid setting for forces.

2. The above explanation indicates that  ADE does not describe the acoustic streaming when the streaming
velocity is small. However ADE can be relevant to `intense' regimes of acoustic streaming, including the
generation of secondary vortices, similar to Langmuir circulations \cite{CraikLeib, VladimirovMHD}.

3. Our approach (based on the two-timing method, distinguished limit, and the unusual choice of small
parameter) is different from the mainstream classical  theory of acoustic streaming. It suggests that ADE can
give new results.

4. The presented asymptotic procedure (based on the equation for vorticity) is remarkably simple. However it
has one  disadvantage:  modified pressure in ADE (\ref{eqn-3-bar1-vel}) and (\ref{eqn-3-bar1-vel-comp})
remains undefined (in terms of physical parameters involved). Therefore the important next step is to repeat
the same asymptotic procedure for the Euler equations, from the very beginning. The expressions for modified
pressure will allow one to calculate the  forces exerted on walls, particles, \emph{etc.}

5. The notion of a drift is actively used in this paper. It is known that a drift velocity can appear from
Lagrangian, Eulerian, or hybrid (Euler-Lagrange) considerations; look for the  discussion related to these
options in \cite{VladimirovMHD}.

6. The relation between high-frequency asymptotic solutions and small-amplitude asymptotic solutions were
studied in \cite{VladimirovX2}; in particular, for  purely periodic non-modulated solutions these classes can
be isomorphic to each other (any solution from one class can be transformed into a solution from another
class).

7. The mathematical justification of the derived averaged equation by the estimation of an error in the
original governing equation can be performed in a similar to
\cite{Vladimirov,VladimirovX2} way.

8. One can also derive the higher approximations of ADE as it has been done in
\cite{Vladimirov,VladimirovX2}. They can be especially useful for the study of motions with
$\overline{\vV}_0\equiv 0$, see \cite{Vladimirov}.

9. Viscosity and diffusivity can be incorporated into our asymptotic theory. However, viscous and diffusion
terms produce additional to (\ref{small-pars-h}) independent small parameters and the result will essentially
depend on their order in terms of $\varepsilon$. In particular, the scale of slow time $s$ (\ref{tau-s-h})
can be different.

10. ADE is valid not only for infinite flows but also for finite flow domains, including the boundaries that
radiates sound.

\begin{acknowledgments}
The author is grateful to Profs. A.D.D.Craik, S.Leibovich, H.K.Moffatt, and N.Riley for helpful discussions.
\end{acknowledgments}

\end{document}